\documentclass[a4paper,10pt]{amsart}

\usepackage{amssymb,epsfig,amsmath,amsthm}
\numberwithin{equation}{section}
\newtheorem{prop}{Proposition}[section]

\begin{document}

\title[Volatility Harvesting]{{\Large Volatility Harvesting: Extracting\\ Return from Randomness}}

\date{November 2015\vspace{0.5cm}\\
Record Currency Management, Windsor, UK}

\author{J. H. Witte}
\address{Record Currency Management, Windsor, UK}
\email{jwitte@recordcm.com}

\begin{abstract}
Studying Binomial and Gaussian return dynamics in discrete time, we show how excess volatility can be traded to create growth. We test our results on real world data to confirm the observed model phenomena while also highlighting implicit risks.\\

\end{abstract}

\maketitle
\setcounter{section}{+1}

\section*{Introduction}

Consider a fair even odds game in which we win or lose a proportion $r$ of our capital in every round. As
\begin{equation}
(1+r)(1-r)=1-r^2,\quad r>0,\label{binf}
\end{equation}
we reduce our capital by factor $r^2$ if we win once and lose once (assuming that every time we stake all our capital). After only two rounds, there is already a $75\%$ probability of being behind, even though the expected value of the game is zero regardless of how many rounds we play. After $M$ rounds of the considered fair game, the median outcome is $(1-r^2)^{M/2}<1$, which is decreasing in $M$.

If we have another (identical but independent) game at our disposal, then we can, in every round, split our capital equally between both games. The question is what we intuitively believe the effect on his wealth progression is going to be, comparing a single game (\emph{imbalanced}) with two simultaneous games (\emph{balanced}).

For the balanced game, outcomes $(1+0.5r)$ and $(1-0.5r)$ have equal probabilities. But, as the returns of two simultaneous games can net to zero, a new neutral state has been added to the player's space of outcomes. We see the effects in Table \ref{tab:binDyn}. After one round of playing two games simultaneously, we obtain less extreme outcomes, but we leave our win-loss ratio unchanged (with a 25\% probability for winning and losing, respectively, and a 50\% probability of a neutral outcome). After two rounds of two simultaneous games, our probability of being behind has been reduced to $43.75\%$, and we have a probability of $25\%$ of breaking even, while the expected value is still zero.

In this simple example, we observe that rebalancing can increase the probability of a positive return. By considering Binomial and Gaussian dynamics, we will now show, with relatively little technical complexity, that the gap in the most likely outcomes of the two strategies continues to widen as time tends to infinity, from which we then infer the rebalancing principle: in the long term, volatility reduction translates into growth.

Throughout the discussion, it is important to bear in mind that trading strategies which generate growth through rebalancing (or volatility harvesting) do require specific market dynamics to persist. They are therefore conceptually no different from a simple directional trade -- we are merely betting on market dynamics rather than market direction, and success is not an arbitrage.

\section*{Binomial Dynamics}\label{Sec_Bin}

We introduce some mathematical notation and relax previous assumptions slightly. Consider two assets, $A_1$ and $A_2$, with returns given by
\begin{equation*}
R_{i,j} = \mu+rB_{i,j},\quad 1\leq j\leq M,\ i=1,\,2,
\end{equation*}
where $M$ denotes the number of time steps considered, and suppose $\mu\in(0,1)$ with $\mu+0.5r=1$ and $\mu-r>0$. Suppose $B_{i,j}\sim \mathcal{B}(1,p)$ are Bernoulli distributed for some parameter $p>0$ and with correlation $\rho=\mathrm{Corr}(B_{1,j},B_{2,j})>0$. Suppose also that there is no serial (inter-) correlation in the considered random processes, such that any random variables drawn at different points in time are independent.

\begin{table}[b]
\caption{\it For $p=0.5$ and $\rho=0$, we see the differences in outcomes between playing a single (imbalanced) or two simultaneous (balanced) games. We notice that the balanced player has a significantly lower probability of falling behind due to the zero return he obtains in a round where he simultaneously wins and loses a game.}
\centering
\begin{tabular}{ c | c | c | c || c}
 & $\mathbb{P}\big[R > 1\big]$ & $\mathbb{P}\big[R = 1\big]$ & $\mathbb{P}\big[R < 1\big]$ & $\mathbb{P}\big[R \geq 1\big]$ \\ \hline\hline
Imbalanced, after $1^{st}$ round &  50\% & 0 & 50\% & 50\%\\ \hline
Balanced, after $1^{st}$ round &  25\% & 50\% & 25\% & 75\%\\ \hline\hline
Imbalanced, after $2^{nd}$ round &  25\% & 0 & 75\% & 25\%\\ \hline
Balanced, after $2^{nd}$ round &  31.25\% & 25\% & 43.75\% & 56.25\%\\ \hline\hline
\end{tabular}
\label{tab:binDyn}
\end{table}

If, at every time $j$, we invest fully into either $A_1$ or $A_2$, we engage in what we earlier termed an imbalanced game, whereas, if we spread our allocation equally between $A_1$ and $A_2$, we engage in a balanced game. The expected period return is identical for imbalanced and balanced strategies, but, for an imbalanced portfolio, the probability of negative period returns is given by 
$$\mathbb{P}\big[R_{i,j} < 1\big] = 1-p,$$
while, for the balanced portfolio, we have
\begin{align}
\mathbb{P}\big[0.5 R_{1,j} + 0.5 R_{2,j} < 1\big]=&\ \mathbb{P}\big[B_{1,j} = 0 \wedge B_{2,j} =0\big]\nonumber\\
=&\ \mathbb{P}\big[B_{1,j}= 0 \mid B_{2,j} =0\big]\,\mathbb{P}\big[ B_{2,j} =0\big]<1-p,\label{FewDEq}\nonumber
\end{align}
which holds regardless of $\rho$.

\begin{figure}[b]
\centering
\includegraphics[scale=0.35]{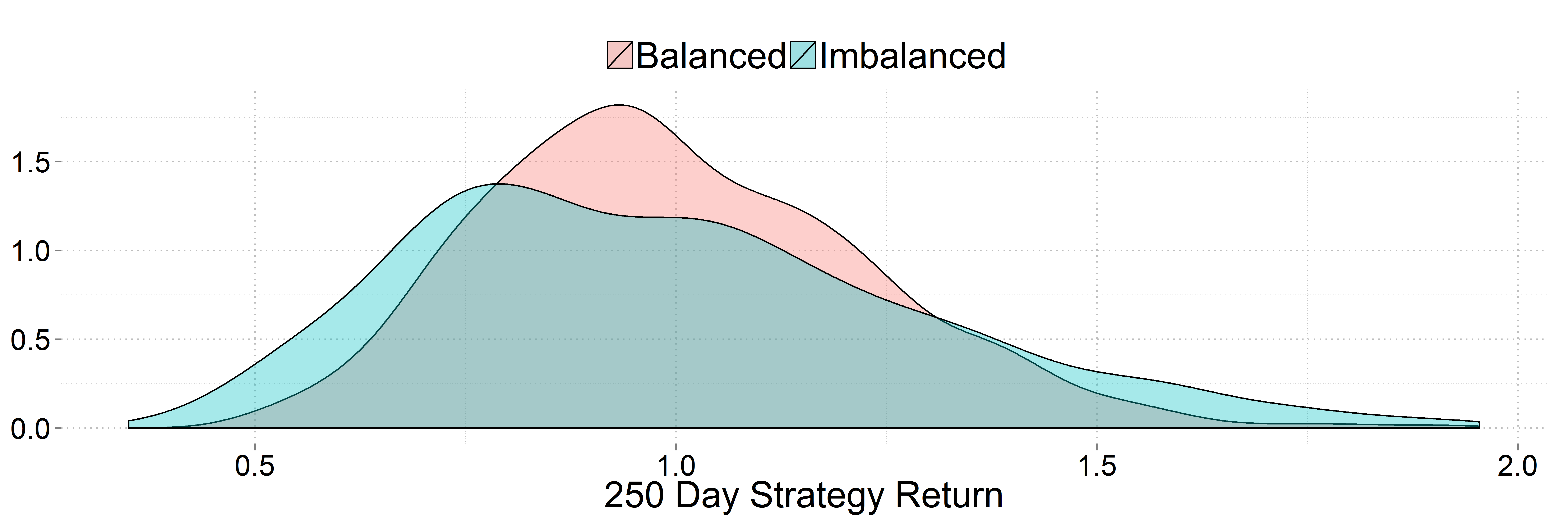}
\caption{The (fitted) densities obtained from 1000 simulated paths of balanced and imbalanced strategies on binomial dynamics. We use $p=0.5$, $\mu=0.98$, $r=0.04$, and $M=250$, which corresponds to one year trading with unbiased daily $\pm 2\%$ dynamics. We use $\rho = 0$, which represents independent assets. The median points of the distributions (which for this choice of $p$ correspond to the most likely outcomes) are $0.9751$ and $0.9512$ for the balanced and imbalanced strategies, respectively.}
\label{BinSample}
\end{figure}

After many rounds of our game, we would expect $pM$ ups and $(1-p)M$ downs. The most likely outcome for the imbalanced strategy is therefore given by
\begin{align*}
\ (\mu+r)^{pM}\mu^{(1-p)M}
= (\mu+r)^{(2p-1)M}(\mu+r)^{(1-p)M}\mu^{(1-p)M}
\end{align*}
Conversely, for the balanced strategy, the most likely outcome is 
\begin{align*}
(\mu+r)^{\beta_{1}M}(\mu+0.5r)^{\beta_{2}M}\mu^{\beta_{3}M}
= (\mu+r)^{(\beta_{1}-\beta_3)M}(\mu+r)^{\beta_3M}\mu^{\beta_{3}M},
\end{align*}
where $\beta_1=\mathbb{P}\big[B_{1,j}=B_{2,j}=1\big]$, $\beta_3=\mathbb{P}\big[B_{1,j}=B_{2,j}=0\big]$, and $\beta_2 = 1-\beta_1-\beta_3$.
We can write
\begin{align*}
\rho =  \frac{\beta_1-p^{2}}{p(1-p)},
\end{align*}
so $\beta_1=p(1-p)\rho+p^2$. By a symmetry argument, we obtain $\beta_3=p(1-p)\rho+(1-p)^2$, and finally $\beta_2=1-\beta_1-\beta_3=2p(1-p)(1-\rho)$. Noting that $\beta_1-\beta_3 = 2p-1$ and $1-p-\beta_3=0.5\beta_2$, we obtain the ratio between the modal values of the imbalanced and balanced strategies as
\begin{align*}
\big[(\mu +r)\mu\big]^{0.5\beta_2M} = \big[\mu + 0.5r\mu\big]^{0.5\beta_2M} < 1
\end{align*}
since $\mu<1$.
We conclude that the distribution of the balanced strategy has a mode (the highest point of the distribution, the value most likely to occur) which is higher than that of the imbalanced strategy, and the gap widens as $M$ increases.

If we consider that the number of outcomes lying near the distribution mode tends to infinity as $M$ increases, then the probability of the balanced strategy outperforming the imbalanced strategy also tends to infinity with $M$. At the same time, expected value of both strategies is always given by $(\mu+rp)^M$.

In Figure \ref{BinSample}, we see (fitted) densities obtained from 1000 simulated paths of balanced and imbalanced strategies on binomial dynamics. We use $p=0.5$, $\mu=0.98$, $r=0.04$, and $M=250$, $\rho = 0$, which corresponds to one year trading with unbiased daily $\pm 2\%$ dynamics. We use $\rho = 0$. The median points of the distributions (which for this choice of $p$ correspond to the most likely outcomes) are $0.9751$ and $0.9512$ for the balanced and imbalanced strategies, respectively, confirming our theoretical results.

\section*{Gaussian Dynamics}\label{Sec_Normal}

Consider two assets $A_1$ and $A_2$ with returns given by
\begin{equation}
R_{i,j} = \mu+\sigma X_{i,j},\quad 1\leq j\leq M,\ i=1,\,2,\label{NormProc}
\end{equation}
where $X^i_j\sim N(0,1)$ are normally distributed with $\mathrm{Corr}(X_{1,j},X_{2,j})=\rho$. As before, we assume that any two random variables drawn at different points in time are independent.
We also suppose $A_{1,0}=A_{2,0}=1$.

We denote by $P$ a portfolio for which, at the beginning of every time period, all capital is split equally between assets $A_1$ and $A_2$. The period returns of this rebalanced portfolio $P$ are given by
\begin{align*}
R_{P,j}= \frac{1}{2}R_{1,j} + \frac{1}{2}R_{2,j}
=\mu + \frac{1}{2}\sigma X_{1,j} + \frac{1}{2}\sigma X_{2,j}.
\end{align*}
The portfolio value $P_M$ at time $M$ is then given by
\begin{align*}
P_M=\prod^M_{j=1}R_{P,j}
\end{align*}
if we set our starting capital equal to 1. We obtain the expected logarithmic growth rate as
\begin{align}
\mathbb{E}\Big[\log P_M^{1/M}\Big] =\frac{1}{M}\sum^M_{j=1}\log R_{P,j}
= \mathbb{E}\log\Big[\mu + \frac{1}{2}\sigma X_1 + \frac{1}{2}\sigma X_2\Big],\label{balancedLogr}
\end{align}
where $X_1$, $X_2\sim N(0,1)$ with $\mathrm{Corr}(X_1,X_2)=\rho$. The expected logarithmic growth rate of the individual assets $A_1$ and $A_2$ is given by
\begin{equation}
\mathbb{E}\Big[\log \left(A_{1,M}\right)^{1/M}\Big] = \mathbb{E}\Big[\log \left(A_{2,M}\right)^{1/M}\Big] = \mathbb{E} \log\Big[\mu + \sigma X^1\Big].\label{imbalancedLogr}
\end{equation}

\subsubsection*{Comparing Growth Rates}

Developing expressions \eqref{balancedLogr} and \eqref{imbalancedLogr} in a second order Taylor expansion about zero, we obtain
\begin{equation*}
\mathbb{E}\Big[\log P_M^{1/M}\Big] = \mu - \frac{1}{2}\sigma^2 + \frac{1}{4}\sigma^2(1-\rho)
\end{equation*}
and
\begin{equation*}
\mathbb{E}\Big[\log \left(A^1_M\right)^{1/M}\Big] = \mu - \frac{1}{2}\sigma^2
\end{equation*}
if we assume that $\sigma>>\mu$. The rebalancing profit in terms of growth rate differential is then given by
\begin{equation}
\frac{1}{4}\sigma^2(1-\rho).\label{RebBonusHeuristic}
\end{equation}
We observe that the rebalancing profit scales inversely with $\rho$. And, for $\rho=1$, the rebalancing profit is zero (as we would expect).
Relying on a more detailed outline given by Breiman (1961), we can now formulate the following result.

\begin{prop}\label{RebBonusTh}
Denote by $\Lambda_{{\rho_1},M}$ and $\Lambda_{{\rho_2},M}$ the time $M$ values of two balanced strategies which differ only in the correlations $\rho_1$ and $\rho_2$, $\rho_1<\rho_2$, of their respectively traded asset pairs, with all other dynamics being equal. We have
\begin{equation*}
\lim_{M\to\infty}\frac{\Lambda_{\rho_1,M}}{\Lambda_{\rho_2,M}}=\infty\quad\text{a.s.}\label{RebBonus_Eq1}
\end{equation*}
for the two strategies. The special case of an imbalanced strategy is contained by chosing $\rho_2=1$, which tells us that, in the long run, the balanced strategy will almost surely outperform the imbalanced strategy obtained by trading only one of the two assets.
\end{prop}

We can generalise the Gaussian dynamics in \eqref{NormProc} by considering
\begin{equation*}
R_{i,j} = \mu_i+\sigma_i X_{i,j}
\end{equation*}
for $\mu_i$ and $\sigma_i$ depending on $i=1$, $2$. Then, denoting by $\theta\in [0,1]$ the portfolio balance of the two assets,
\begin{align*}
R_{P,j}=&\ \theta R_{1,j} + (1-\theta)R_{2,j}\\
=&\ \theta\mu_1+(1-\theta)\mu_2 + \theta\sigma_1 X_{1,j} + (1-\theta)\sigma_2 X_{2,j},
\end{align*}
and so, developing this expression as before, we obtain
\begin{align}
\mathbb{E}\Big[\log P_M^{1/M}\Big] =&\ \theta\mu_1+(1-\theta)\mu_2
-\frac{1}{2}\Big[\theta^2\sigma_1^2+(1-\theta)^2\sigma_2^2\Big]-\theta(1-\theta)\sigma_1\sigma_2\rho,\label{GenEq}
\end{align}
while
\begin{equation*}
\mathbb{E}\Big[\log \left(A_{i,M}\right)^{1/M}\Big] = \mu_i-\frac{1}{2}\sigma_i^2.
\end{equation*}

In many cases, a careful choice of $\theta$ can be used to create an expected logarithmic growth rate for the balanced strategy which exceeds that of both individual assets, and the result of Proposition \ref{RebBonusTh} extends to those situations.
In particular, a positive expected logarithmic growth rate for the balanced strategy can be achieved even in some situations where both assets individually have negative expected logarithmic growth.

\section*{General Market Dynamics}

Provided $X_1$ and $X_2$ are identically distributed, we can apply Jensen's inequality to conclude directly that
\begin{align}
\mathbb{E}\log\Big[\theta X_1 + (1-\theta)X_2\Big]\geq\ \theta\,\mathbb{E}\log X_1 + (1-\theta)\,\mathbb{E}\log X_2 = \mathbb{E}\log X_1 \label{JenIneq}
\end{align}
without specifying the probabilistic dynamics of our two traded assets any further.

Intuitively, \eqref{JenIneq} can be used to explain why the results of Proposition \ref{RebBonusTh} would be expected to hold more generally, and why normality is a sufficient but not necessary prerequisite for the presented results. Caution is required to ensure that the limit $M\to\infty$ can be taken safely, and the Gaussian dynamics as studied in the previous section allow the use of Breiman's (1961) classic results.
Details of a comprehensive and general proof, which requires more technicality, can be found in \cite{JoyOfVol}.

\section*{Relationship to Kelly's Formula}

The Kelly criterion \cite{KellyIR} can be stated as maximising the expected logarithmic growth rate under certain conditions. For two assets in a multi-period model, the optimal weights $w^*_1$ and $w^*_2$ to invest in asset one and two, respectively, are given by
\begin{equation}
(w^*_1,w^*_2)^T = \Sigma^{-1}(\mu_1,\mu_2)^T,\label{KellyEq1}
\end{equation}
where $\Sigma$ denotes the $2\times 2$ covariance matrix and $\mu_1$, $\mu_2>0$ denote the expected period returns of our two assets. (We require $\Sigma$ to be invertible.) If we denote the  determinant of $\Sigma$ by $|\Sigma|$, $|\Sigma|>0$, then we can write \eqref{KellyEq1} as
\begin{equation*}
(w^*_1,w^*_2)^T = 
\frac{1}{|\Sigma|}
\left( \begin{array}{cc}
\sigma^2_2 & \text{-Cov}(X_1,X_2) \\
\text{-Cov}(X_1,X_2) & \sigma^2_1 \\
\end{array} \right)
\left( \begin{array}{c}
\mu_1\\
\mu_2 \\
\end{array} \right),
\end{equation*}
which tells us that, as long as $\mu_1=\mu_2$ and $\sigma_1=\sigma_2$, Kelly always commands a balanced strategy with equal amounts invested in assets $A_1$ and $A_2$. (Note that this does not mean that Kelly commands a 50\%-50\% allocation.)

While our previous results state that a higher rebalancing speed will guarantee outperformance of otherwise equal strategies in the long run, Kelly's allocation is one that guarantees outperformance of any other strategy in the long run \cite{Breimann}. Given an imbalanced allocation, it is therefore instructive to view a balanced strategy as an improvement step towards Kelly's allocation, the practical difficulty of the latter being the need for a precise understanding of market dynamics.

\begin{figure}[b]
  \centering
    \includegraphics[width=1\textwidth]{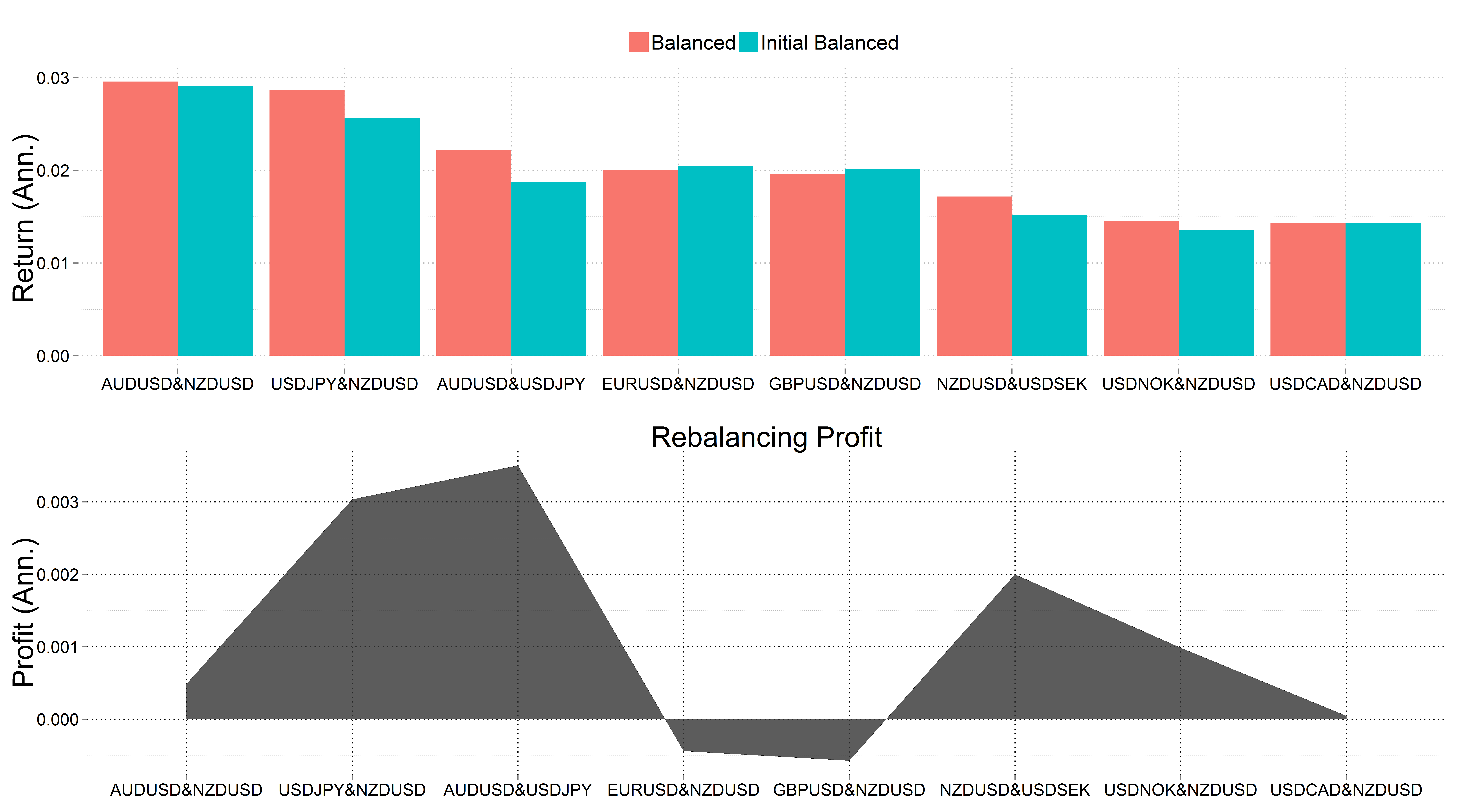}
 \caption{Rebalancing effect for two asset portfolios in G10, 1 of 5.}
\label{g10Reb1}
\end{figure}

\begin{figure}[b]
  \centering
    \includegraphics[width=1\textwidth]{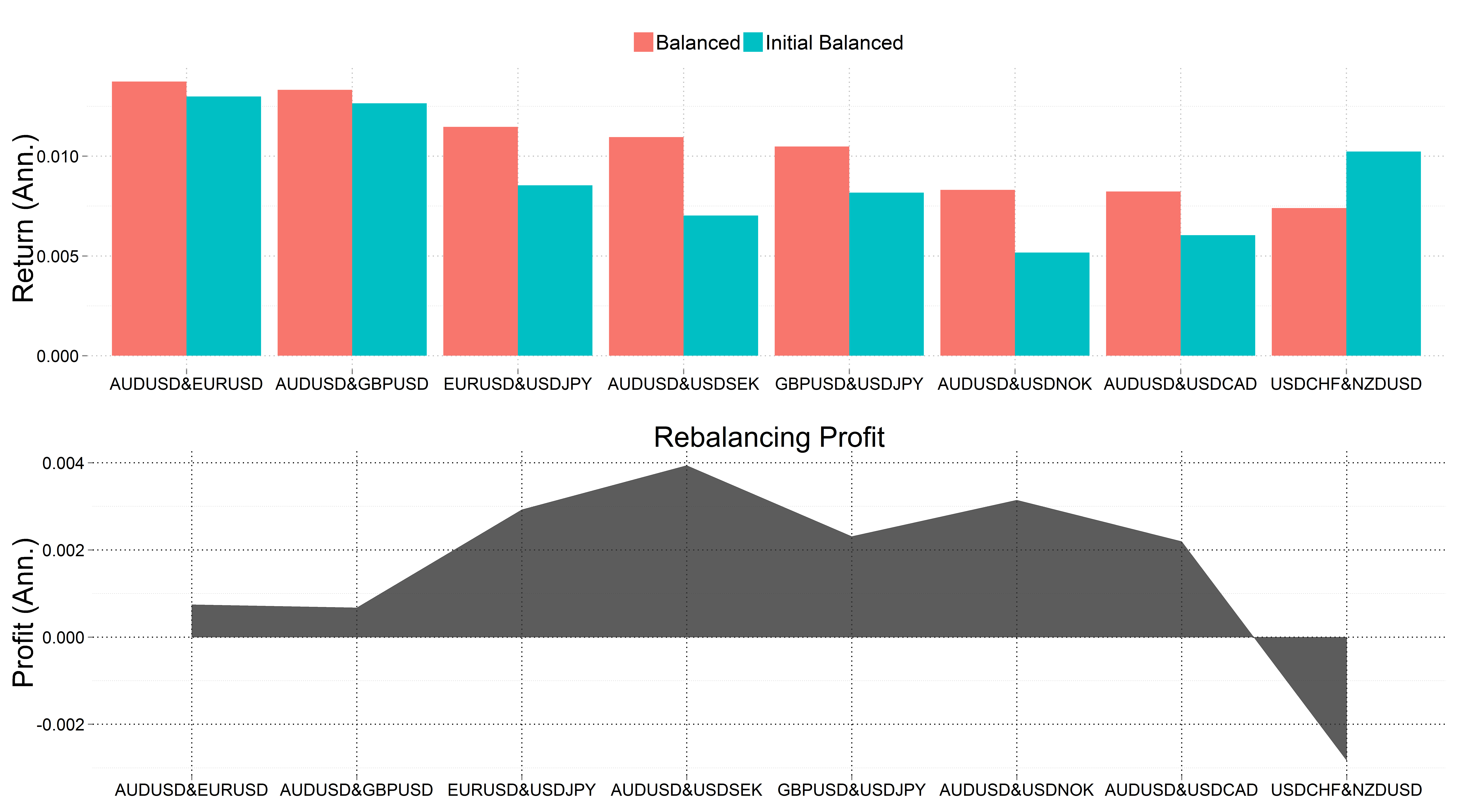}
 \caption{Rebalancing effect for two asset portfolios in G10, 2 of 5.}
\label{g10Reb2}
\end{figure}

\begin{figure}[b]
  \centering
    \includegraphics[width=1\textwidth]{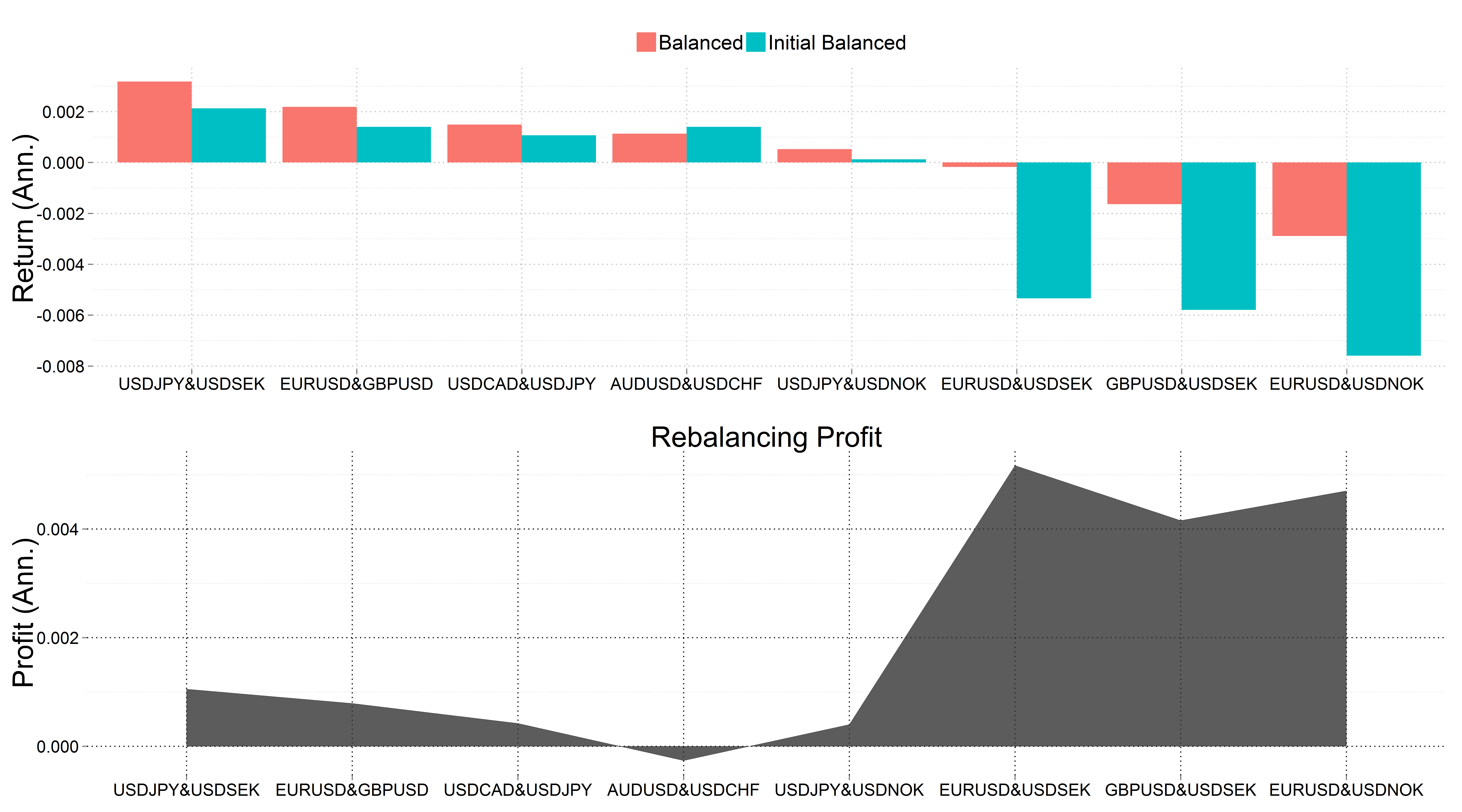}
 \caption{Rebalancing effect for two asset portfolios in G10, 3 of 5.}
\label{g10Reb3}
\end{figure}

\section*{Real World Data}

We use daily WM/Reuters FX data from 1 January 2000 to 5 November 2015 for all G10 USD crosses. We assume the role of a USD based investor. To account for interest rate as well as spot movements, we simulate trading based on historical mid prices of one day forward contracts with gains and losses reconverted into the base currency at the end of every day. We trade such that, for every currency cross, we go longt the first currency and short the second.

For every combination of two exchange rates, we create a two asset portfolio, where we compare the performance of a daily balanced two asset portfolio to that of an imbalanced one (with a 50\%--50\% initial allocation) which we refer to as \emph{initial balanced} portfolio.

\begin{figure}[b]
  \centering
    \includegraphics[width=1\textwidth]{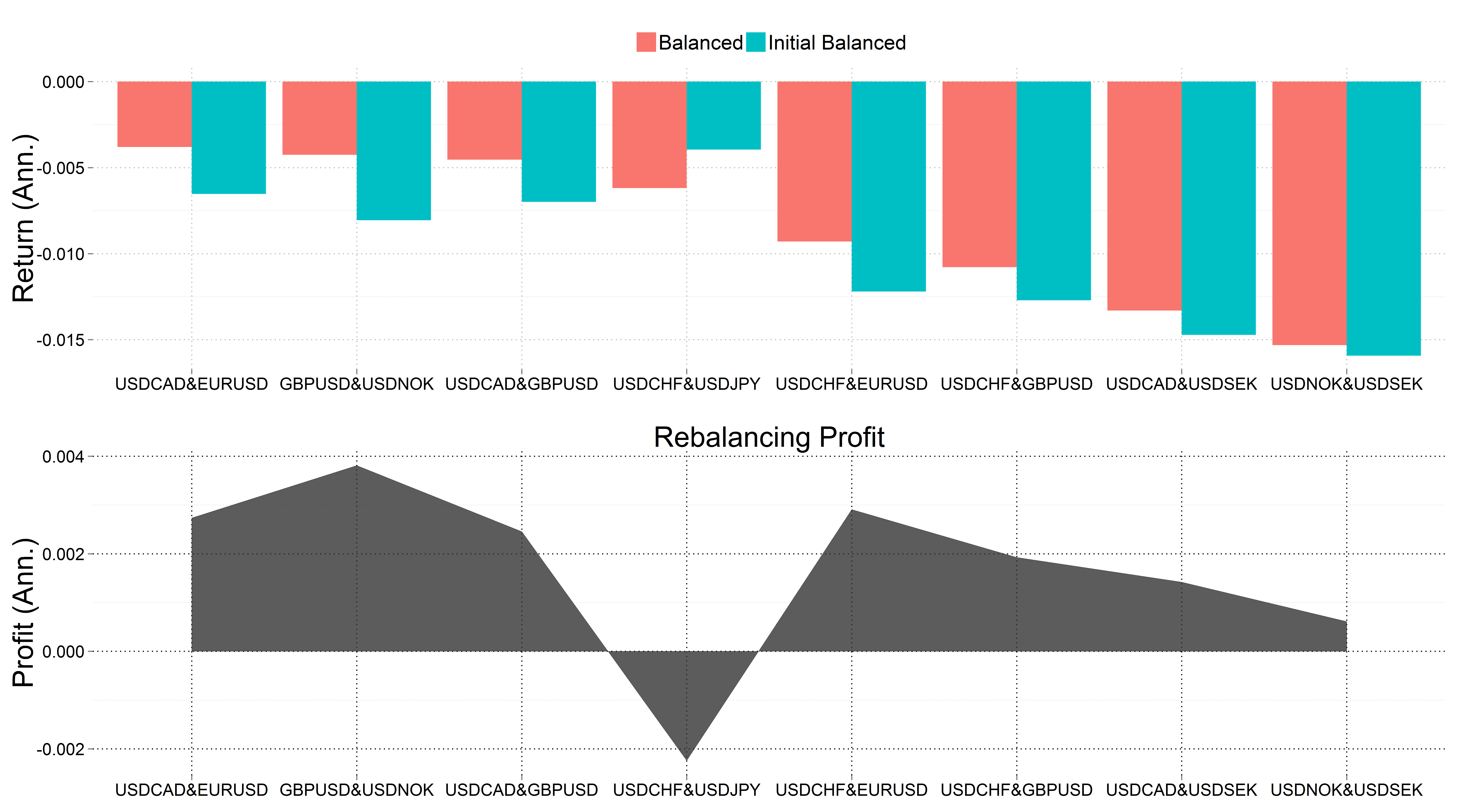}
 \caption{Rebalancing effect for two asset portfolios in G10, 4 of 5.}
\label{g10Reb4}
\end{figure}

\begin{figure}[b]
  \centering
    \includegraphics[width=1\textwidth]{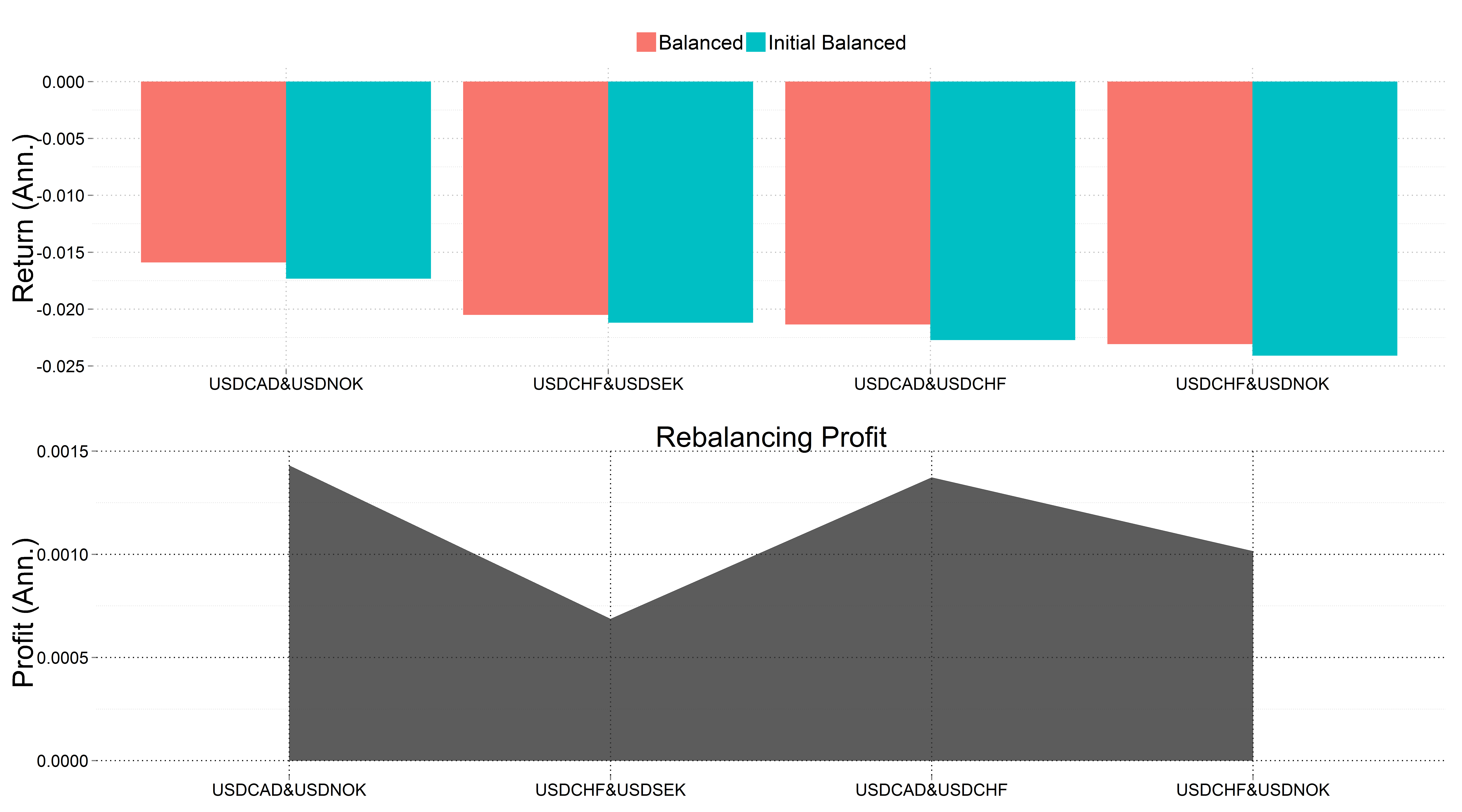}
 \caption{Rebalancing effect for two asset portfolios in G10, 5 of 5.}
\label{g10Reb5}
\end{figure}

In Figures \ref{g10Reb1} to \ref{g10Reb5}, we see the annualised returns generated by the balanced and initial balanced strategies, respectively, as well as the annualised return differential. We notice that, for30 out of the 36 crosses, the rebalanced return is greater than the initial balanced return, albeit by very different magnitudes.

Considering a daily volatility of $0.75\%$ per currency pair and trading on 250 business days, the total rebalancing cost every day would roughly be $2\times 250\times 0.0075 \times \text{half-spread}$, which, for example, gives annual costs of 3.75 bps and 9.375 bps for assumed spreads of 1 bps and 5 bps, respectively; a calculation which does not yet account for the (relatively cheaper) cost of rolling the underlying position, and which does not yet take into account occasional big movements. If we consider 20 bps as a total annual cost number for very currency liquid pairs and highly efficient execution, a net annual profit of 10 or 20 bps seem feasible for some currency pairs. For many currency pairs, the potential rebalancing profit will in the same order of magnitude as the required transaction costs.

\section*{Conclusion}

Showing that exponential growth can be generated in a random market with growth rate zero is an idea originally presented by Shannon (cf.\,\cite{Poundstone}). For $\frac{1}{2}\sigma_1^2\geq\mu_1>\frac{1}{4}\sigma_1^2$, $\mu_2=0$, $\sigma_2=0$, $\rho=0$, and $\theta=0.5$, we obtain the logarithmic growth rate
\begin{align*}
\mathbb{E}\log P_M^{1/M} = \frac{1}{2}\mu_1 - \frac{1}{8}\sigma^2_1 >0
\end{align*}
in \eqref{GenEq} for a strategy which rebalances between an asset with non-positive growth rate and an interest free cash account, and we recover Shannon's strategy from our results. In analogy with Proposition \ref{RebBonusTh}, we have no finite time arbitrage (and therefore no conflict with the fundamental theorem of asset pricing), but we can expect profitability in the long run.

Is it surprising, or even contradictory, that exponential growth can be generated from volatility in a zero-growth (but otherwise random) market?
Not if we recall that the balanced strategy has a smaller probability of getting lucky (i.e., of profiting from random big moves), and that expected values remain unchanged, as is for example highlighted in Figure \ref{BinSample}.

In practice, while rebalancing is not confined to mean-reversion environments, it still relies on continuity of dynamics, which constitutes the strategy's risk, namely that any observed large deviation will provoke doubt as to whether the required equilibrium in the underlying assets is still being assessed correctly. This need for repeated correct assessment of the market environment highlights a property volatility harvesting has in common with many other strategies: success depends on skilful application.

\bigskip
{\bf DISCLAIMER}

The views expressed are those of the author and do not reflect the official policy or position of Record Currency Management.

\end{document}